\documentclass[twoside,web]{ieeecolor}
\usepackage{generic}
\usepackage{cite}
\usepackage{amsmath,amssymb,amsfonts}
\usepackage{algorithmic}
\usepackage{graphicx}
\usepackage[colorlinks,linkcolor=.,anchorcolor=., citecolor=.,filecolor=.,menucolor=., runcolor=., urlcolor=blue]{hyperref}
\usepackage{float}
\usepackage{multirow}
\usepackage{footnote}
\usepackage{hyperref}
\hypersetup{
    colorlinks,
    linkcolor={blue!50!black},
    citecolor={blue!50!black},
    urlcolor={blue!80!black}}
\usepackage{svg}
\makesavenoteenv{tabular}
\makesavenoteenv{table}
\usepackage{caption}
\usepackage{setspace}
\usepackage{subcaption}
\usepackage{hyperref}
\usepackage{verbatim}
\usepackage{textcomp}
\usepackage[capitalize,nameinlink]{cleveref}
\usepackage{tikz}
\def\BibTeX{{\rm B\kern-.05em{\sc i\kern-.025em b}\kern-.08em
    T\kern-.1667em\lower.7ex\hbox{E}\kern-.125emX}}
\begin{document}
\fboxrule=0pt
\title{Band Tail State Broadening in IGZO TFTs After pBTI-Induced Negative $V_\text{T}$ Shift Revealed via DC and 1/f Noise Measurements}
\author{R. Asanovski*, P. Rinaudo*, A. Chasin, Y. Zhao, H.F.W. Dekkers, M. J. van Setten, D. Matsubayashi, \\ N. Rassoul, A. Belmonte, G.S. Kar, B. Kaczer, J. Franco \vspace*{-0.9cm}
\thanks{\scriptsize{*Equally contributed to this work.\\
This work has been enabled in part by the NanoIC pilot line. The acquisition and operation are jointly funded by the Chips Joint Undertaking, through the European Union’s Digital Europe (101183266) and Horizon Europe programs (101183277), as well as by the participating states Belgium (Flanders), France, Germany, Finland, Ireland and Romania. For more information, visit nanoic-project.eu. Pietro Rinaudo is supported by a PhD Fellowship of the Research Foundation - Flanders (Belgium) (grant number 1SE2723N).}}}

\maketitle
\begin{abstract}
We investigate the origin of negative threshold voltage shifts in back-gated amorphous IGZO TFTs under positive bias and high temperature stress. Combined DC and 1/f noise measurements reveal that the stress does not generate new dielectric traps but instead {broadens the IGZO conduction band tail states. A recovery experiment confirms that the associated threshold voltage, subthreshold swing, and noise degradation are reversible.} Simulations {using an in-house Poisson solver} confirm the experimental observations that high-temperature stress increases hydrogen doping and {the density of sub-gap states}. 
\end{abstract}

\section{Introduction}
\label{sec:introduction}
Amorphous InGaZnO$_4$ (IGZO) thin-film transistors (TFTs) are promising for future DRAM architectures thanks to their low off-state current and compatibility with large-area, low-temperature processing \cite{Belmonte2021, Hu2019}. Although their electrical performance is well established \cite{Subhechha2021}, their reliability, especially at elevated temperatures, remains challenging \cite{Chasin2021,Wu2022,Chasin2024}. \par
At high temperatures, IGZO TFTs under positive gate bias show complex Bias Temperature Instability (BTI) behavior with two competing mechanisms: electron trapping in gate dielectric defects (causing a positive threshold voltage shift, $\Delta V_{\text{T}} > 0$ V) and hydrogen release from the dielectric, which incorporates into IGZO and acts as a donor, leading to an {\it abnormal negative} shift ($\Delta V_{\text{T}} < 0$ V) \cite{Chasin2021,Liu24}. This negative shift is often ascribed to changes in {energy states near the valence band of the channel}, likely related to IGZO metal-metal bond complexes (M-M) \cite{Wu2022,LAIV_DOS}.\par 
{Previous studies on 1/f noise in IGZO TFTs fabricated in research labs focused on its origin and correlation with factors such as composition \cite{10980071}, channel geometry \cite{10980071,4812043}, radiations \cite{Lee2025}, and hydrogen content at time zero \cite{7544475}. {In contrast, this work exploits the sensitivity of 1/f noise to probe changes associated with the abnormal negative $\Delta V_{\text{T}}$ observed under high‑$T$ gate voltage stress, a critical issue for DRAM applications \cite{Chasin2021,Wu2022,Chasin2024}. Moreover, we study back-gated IGZO TFTs fabricated in a 300mm CMOS-compatible line, exhibiting quasi-ideal subthreshold swing (SS) and minimal threshold voltage variability. By combining DC measurements, 1/f noise, and physics-based simulations, we clearly demonstrate a link between negative $\Delta V_{\text{T}}$ and stress‑induced modifications of the IGZO sub‑gap states.}}

\section{Results and discussion}
\label{sec:results}
\begin{figure}[!t]
\centering
\includegraphics[width=0.23\textwidth]{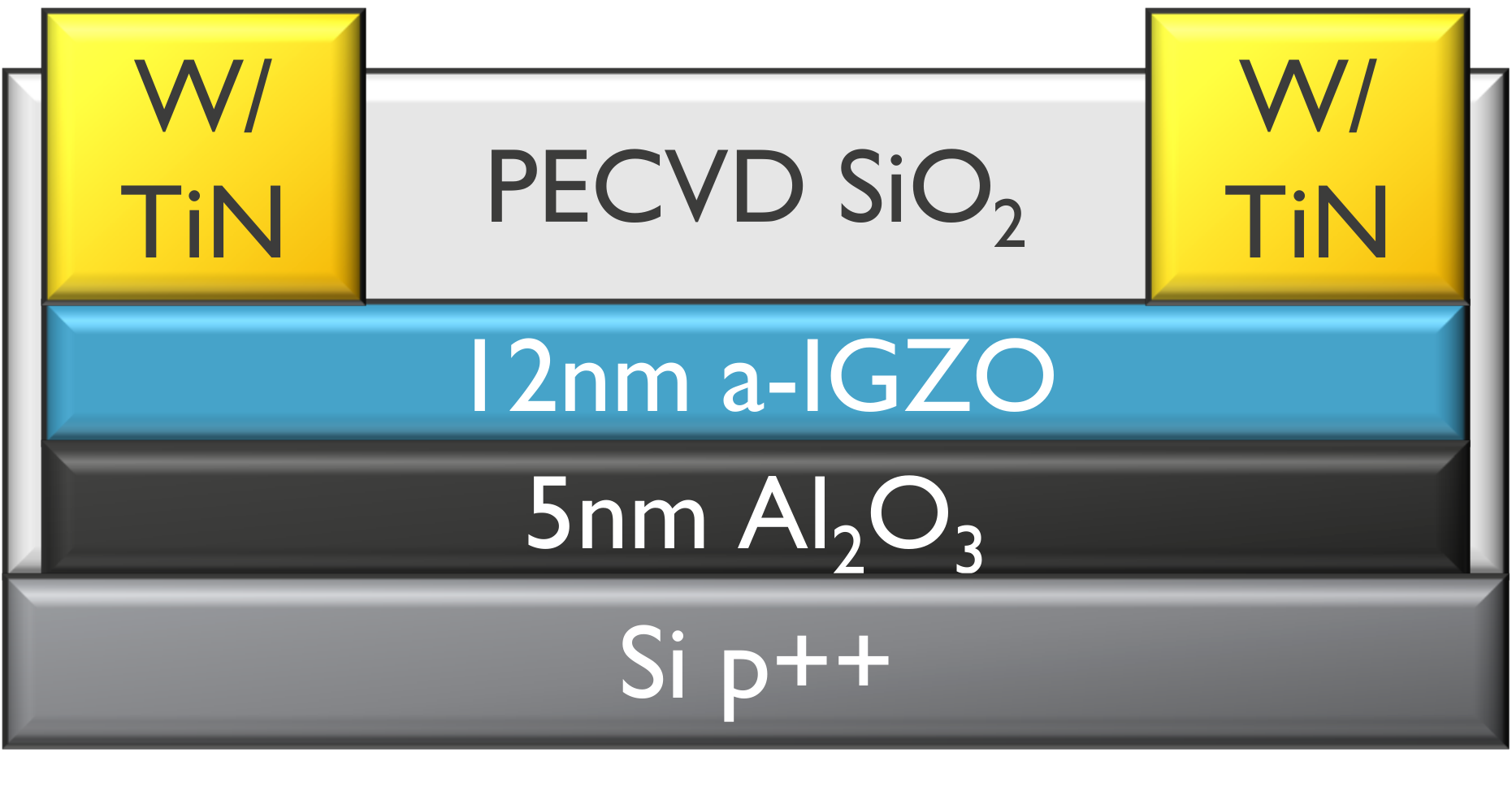} 
\caption{Schematic of the back-gated IGZO TFTs used in this study.}
\label{fig:Schematic}
\end{figure}

We studied large area ($W$=$L$=10 \textmu m) back-gated IGZO TFTs fabricated on a 300 mm wafer to minimize device-to-device variations. The gate consists of heavily doped silicon (p++), with a 5 nm Al\textsubscript{2}O\textsubscript{3} dielectric deposited by atomic layer deposition (ALD) { with TMAH as precursor at 300~°C (\cref{fig:Schematic}). The 12 nm IGZO channel is deposited by PVD under pulsed-DC method from a single target 1:1:1 at room temperature at 100\% Ar concentration. XRF measurements indicate an atomic concentration of In36\%, Ga40\% and Zn24\%.  TiN interlayer and W are deposited as S/D contacts, and PECVD SiO\textsubscript{2} is used as encapsulation layer. Finally, the device is annealed at 350°~C for one hour in oxygen environment. The atomic concentration of hydrogen in as-deposited IGZO film measured by ERD is 0.8\%.} First, we recorded DC I-V characteristics and 1/f noise of the fresh TFTs at $T$= 25~°C. Then, we stressed multiple devices at $T$=125 °C for 900 s at increasing gate voltages, 
inducing progressively higher negative $\Delta V_{\mathrm{T}}$. Each device was stressed only once at a specific $V_{\mathrm{GS,stress}}$. After stressing each device, we cooled them to $T$= 25 °C to characterize DC and noise. This minimized $V_{\mathrm{T}}$ recovery, as it is a thermally activated process \cite{Chasin2021}, and avoided measurement drifts during the following noise characterization. $V_{\mathrm{T}}$ was extracted at $I_{\mathrm{D}}=$100~nA, and the SS as the median over $I_{\mathrm{D}}$ from 0.5~nA to 5~nA. {All measurements were done with a Keysight B1500 semiconductor parameter analyzer and a Keysight E4727B low-frequency noise analyzer}. 
\begin{figure}[!tb]
    \centering
    \begin{subfigure}[t]{0.5\columnwidth}
        \centering
        \begin{tikzpicture}
            \node[anchor=south west,inner sep=0] (image) at (0,0) {\includegraphics[width=1\columnwidth]{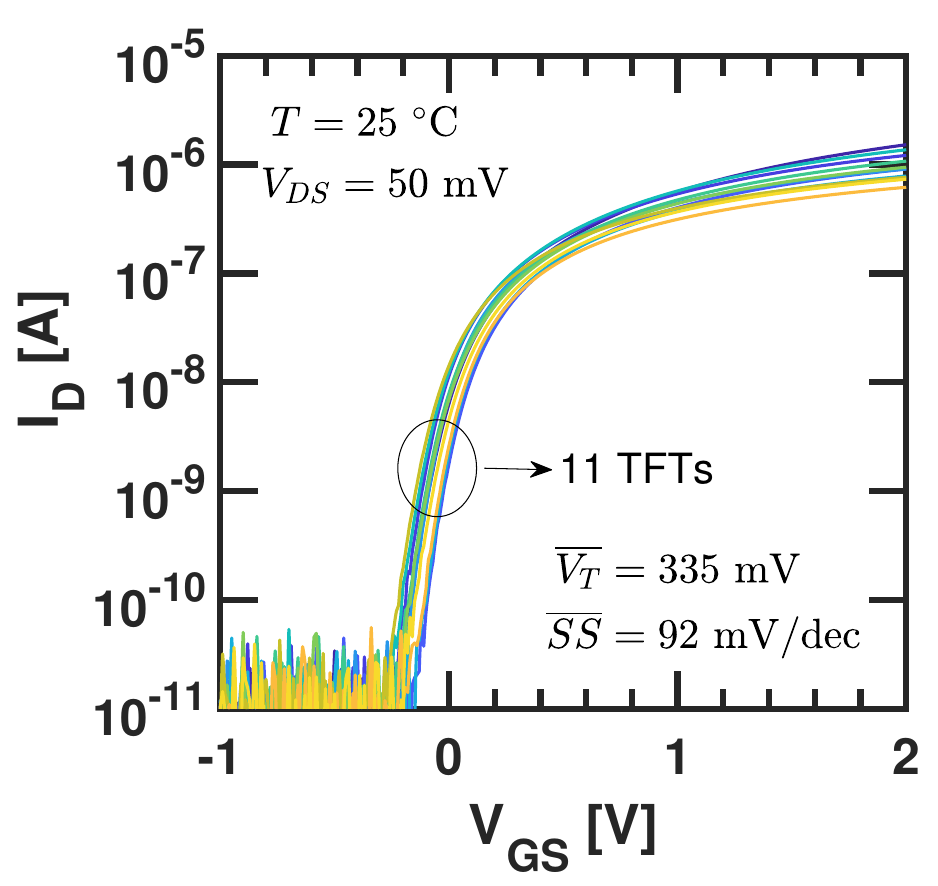}};         
            \node[align=left, anchor=north west] at (1.2,1.7) {(a)};
        \end{tikzpicture}
        \phantomsubcaption

        \label{fig:IdVg_ALL}
    \end{subfigure}%
    \begin{subfigure}[t]{0.5\columnwidth}
        \centering
        \begin{tikzpicture}
            \node[anchor=south west,inner sep=0] (image) at (0,0) {\includegraphics[width=1\columnwidth]{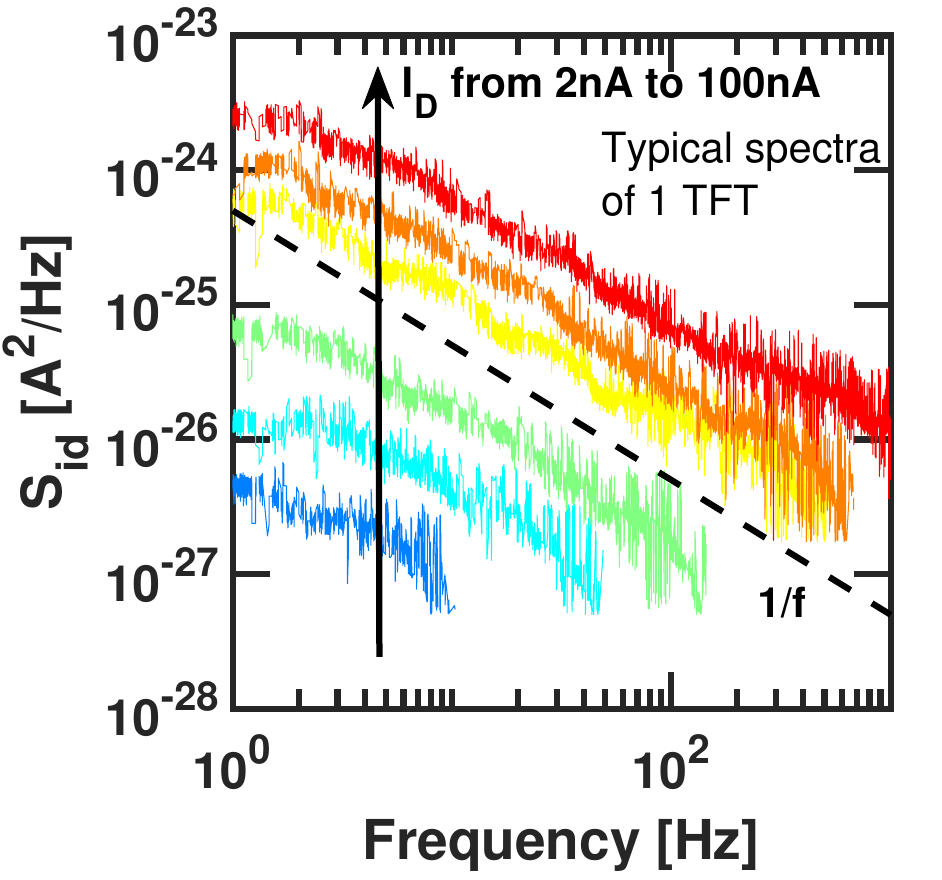}}; 
            \node[align=left, anchor=north west] at (1.1,1.45) {(b)}; 
        \end{tikzpicture}
        \phantomsubcaption

        \label{fig:Sid}
    \end{subfigure}
    \caption{(a) $I_{\textrm{D}}{-}V_{\textrm{GS}}$ and (b) $S_{\textrm{id}}$ vs frequency in the linear region for the fresh IGZO TFTs described in the text.}
    \label{fig:TimeZero_DCNoise}
\end{figure}
\par \cref{fig:TimeZero_DCNoise} shows typical $I_{\mathrm{D}}-V_{\mathrm{GS}}$ and drain current noise power spectral densities (PSDs) in the linear region for the fresh IGZO TFTs used in this study. We interpret the normalized drain current noise PSDs ($S_{\mathrm{id}}/I_D^2$) in subthreshold ($I_D$ from 2~nA up to 100~nA) using the carrier number fluctuation (CNF) model with correlated mobility fluctuations (MF) \cite{Ghibaudo1991}
\begin{equation}
\frac{S_{\mathrm{id}}}{I_D^2}=\frac{qkTN_{\mathrm{BT}}}{WLC_{\mathrm{ox}}^2\alpha}\cdot \frac{1}{f}\cdot \left(\frac{g_{\mathrm{m}}}{I_D}\right)^2\cdot\left(1+{\Omega} \frac{I_{\mathrm{D}}}{g_{\mathrm{m}}}\right)^2\ \ \ ,
\label{eq:SvgSimple}
\end{equation}
{where $g_{\mathrm{\mathrm{m}}}$ is the transconductance, $N_{\mathrm{\textrm{BT}}}$ is an effective trap density, $\alpha$ is a tunneling coefficient estimated with the {Wentzel–Kramers–Brillouin (WKB)} approximation, $ C_{\mathrm{\mathrm{ox}}}$ is the gate capacitance per unit area, $q$ is the elementary charge, $k$ is the Boltzmann constant, $T$ is the temperature, $f$ is the frequency, and $\Omega$ is a parameter quantifying MF.
\par We fit each noise spectrum with a 1/f curve and plot $S_{\mathrm{id}}\cdot\nobreak f/I_D^2$ versus $I_{\mathrm{D}}$ to determine whether the time-zero noise data falls in the CNF or MF region. \cref{fig:Sid_Id_MF} shows that the noise data fits in a MF model because the $S_{\mathrm{id}}\cdot f/I_D^2$ is independent on $I_D$ (see Eq. \ref{eq:SvgSimple}). 
In silicon MOSFETs, noise in the subthreshold region is typically described by the CNF theory, which links 1/f noise to fluctuations of dielectric charge \cite{Christensson68,Ghibaudo1989}. On the other hand, noise in the MF region is often associated with disorder in the channel \cite{Hooge69,Vandamme94}, suggesting that 1/f noise in IGZO TFTs probes the fluctuation of energy states in the channel rather than in the dielectric.
\begin{figure}[]
    \centering
    \includegraphics[width=0.44\textwidth]{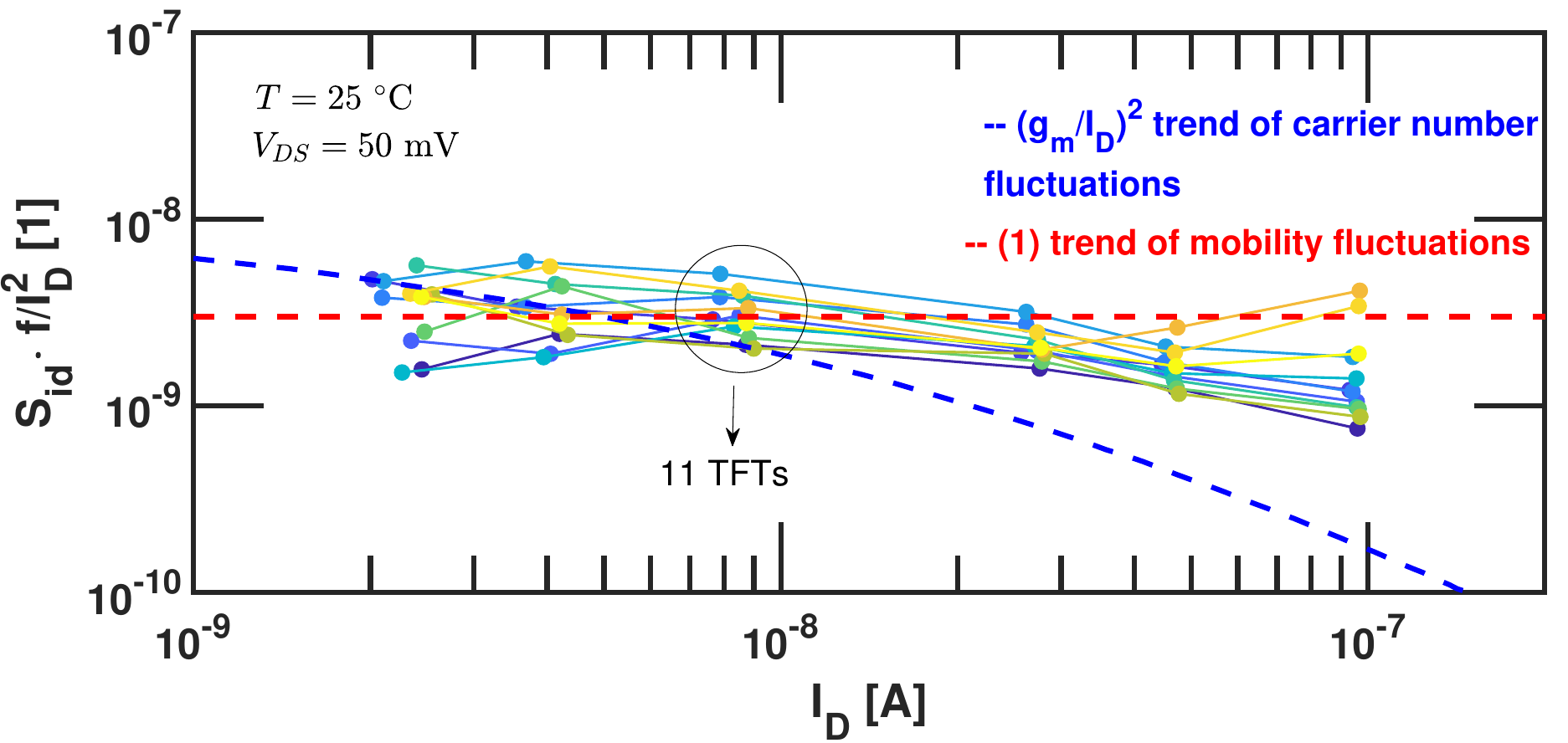}
    \caption{Plot of $S_{\mathrm{id}}\cdot f/I_D^2$ vs $I_{\mathrm{D}}$ identifying MF as the main mechanism for 1/f noise in the fresh IGZO TFTs described in the text.}
    \label{fig:Sid_Id_MF}
\end{figure}
\par \cref{fig:IdVg_Stressed} compares the $I_{\mathrm{D}}-V_{\mathrm{GS}}$ in the linear region before and after stress for all the stress conditions considered in this study. The negative $\Delta V_{\mathrm{T}}$ increases with increasing stress; notably, the SS gradually degrades as well. \cref{fig:Noise_Stressed}a-d shows that the noise in subthreshold increases with increasing stress conditions, \textit{similar to the SS degradation}, indicating an increase of trap-like states below the conduction band minimum energy. { For the noise analysis, we used a model‑independent approach and focus on the relative change in 1/f noise, since this provides a transparent indicator of stress‑induced degradation and avoids relying on mobility‑fluctuation models that remain debated in Si MOSFETs \cite{Vandamme08, Fleetwood15} and are not physically validated for IGZO TFTs.} \cref{fig:Noise_Stressed}e highlights the linear correlation between SS and noise increase in subthreshold, further suggesting that noise and SS are degrading for the same root cause. Note that the noise near $V_{\mathrm{T}}$ does not degrade (see data near $I_\mathrm{D}{=}100$ nA in Fig. \ref{fig:Noise_Stressed}a-d), indicating that the degradation occurs only below $V_{\textrm{T}}$. 
{\par To check if the stress‑induced degradation is reversible, we performed a recovery experiment on a device previously stressed at $V_{\mathrm{GS,stress}}=2.50$ V and $T=125~^\circ$C for 900 s. The device was allowed to relax at $V_{\mathrm{GS}}=0$ V and $T=125~^\circ$C for one week. As shown in Fig.~\ref{fig:Recovery}, the negative $V_{\mathrm{T}}$ shift, as well as the increases in SS and subthreshold noise, nearly return to their fresh values, demonstrating that the degradation is not permanent. These results indicate that the degradation arises from reversible DoS changes rather than irreversible trap generation in the dielectric or in the channel.}
 \begin{figure}[!tb]
\centering
\includegraphics[width=0.44\textwidth]{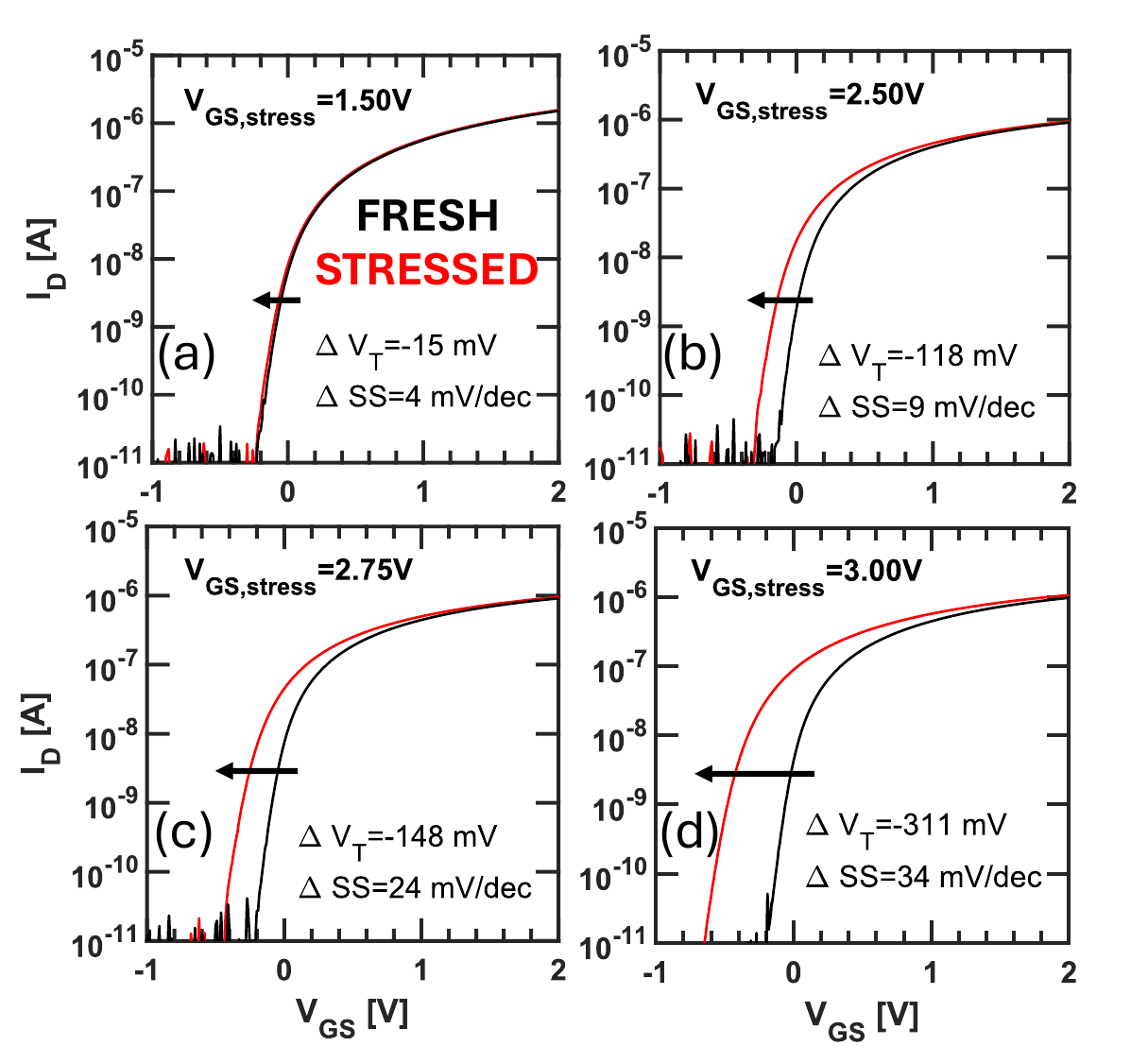} 
\caption{Comparison of $I_{\textrm{D}}{-}V_{\textrm{GS}}$ between fresh TFTs and those stressed at $T$=125~°C with different $V_{\mathrm{GS,stress}}$. The measurements are done at $T$=25~°C and $V_{\mathrm{DS}}$=50~mV. }
\label{fig:IdVg_Stressed}
\end{figure}
\begin{figure}[!tb]
    \centering
    \includegraphics[width=0.36\textwidth]{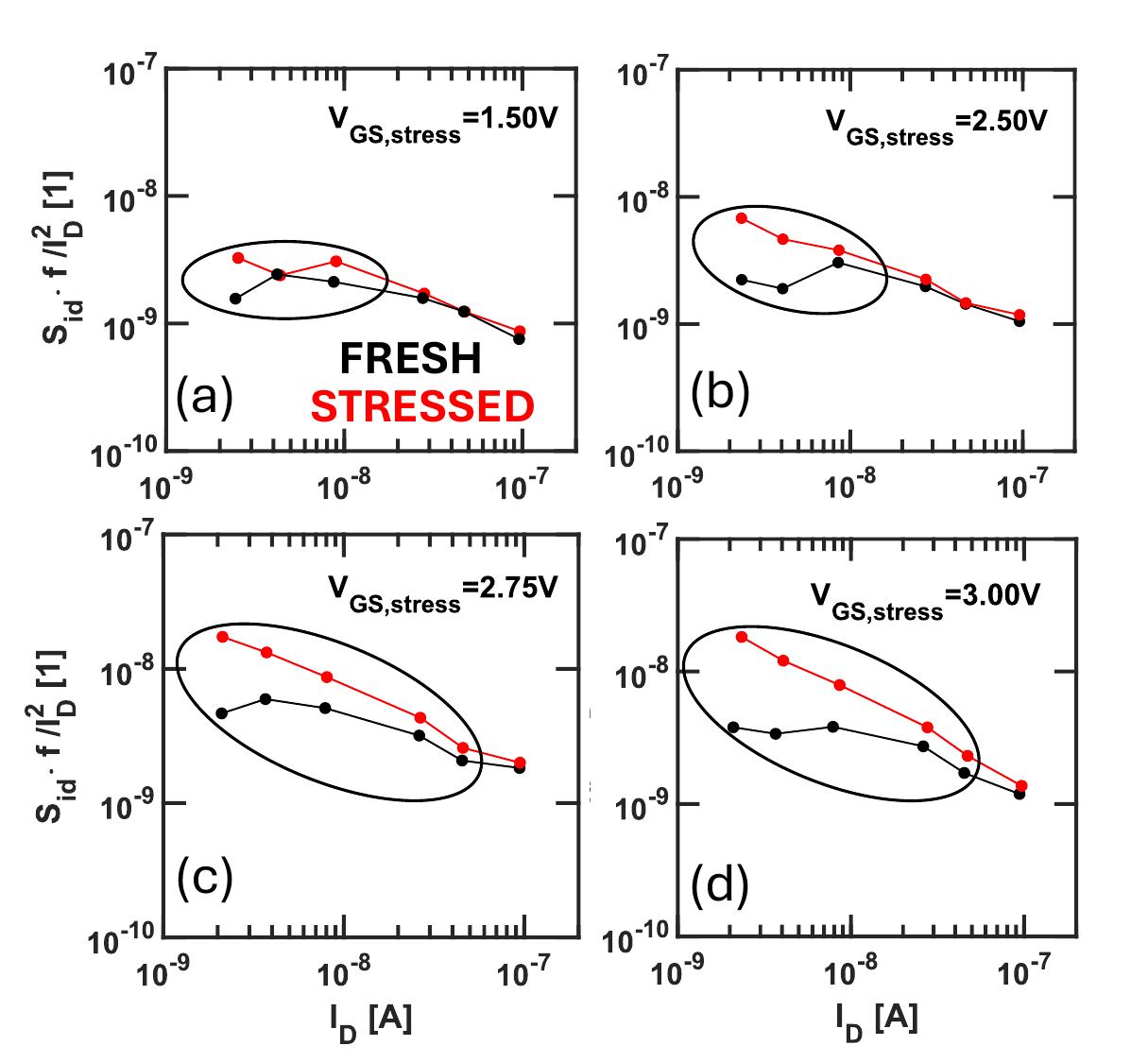}
    \centering
    \includegraphics[width=0.12\textwidth]{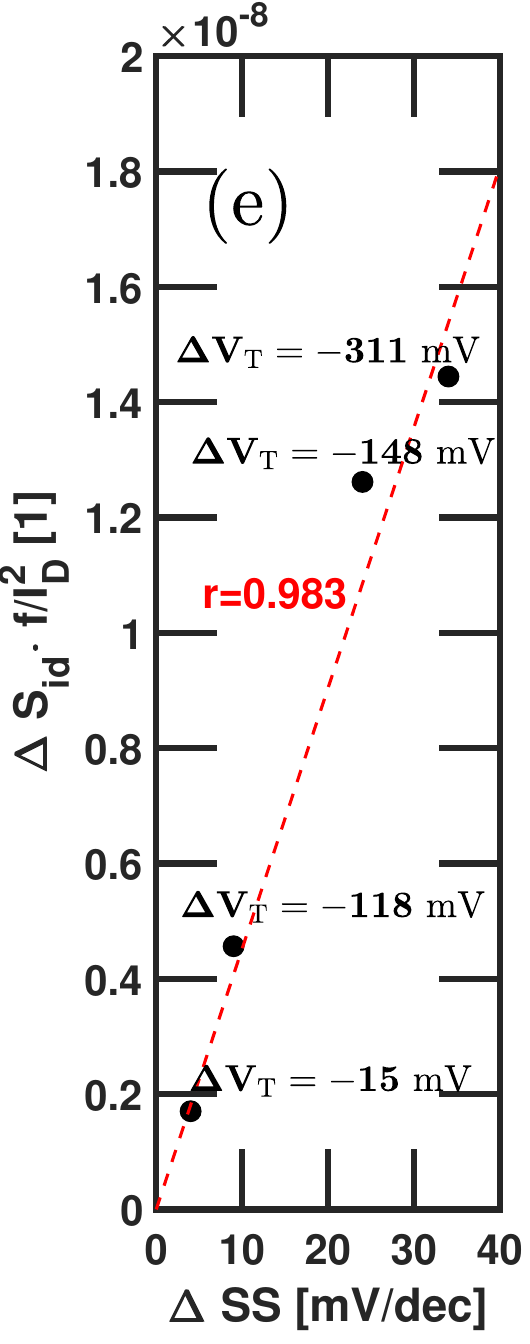} \\ 
    \caption{(a)-(d) Comparison of 1/f noise (plotted as $S_{\mathrm{id}}\cdot f/I_D^2$) between fresh TFTs and those stressed at $T$=125 °C with different $V_{\mathrm{GS,stress}}$. The measurements are done at $T$=25 °C and $V_{\mathrm{DS}}$=50 mV. (e) Linear correlation plot between the increase of SS and noise (at $I_{\mathrm{D}}=$2 nA) for the different negative $\Delta V_{\mathrm{T}}$ induced after PBTI stress.}
    \label{fig:Noise_Stressed}
\end{figure}

\begin{figure}[!tb]
    \centering
    \includegraphics[width=0.24\textwidth]{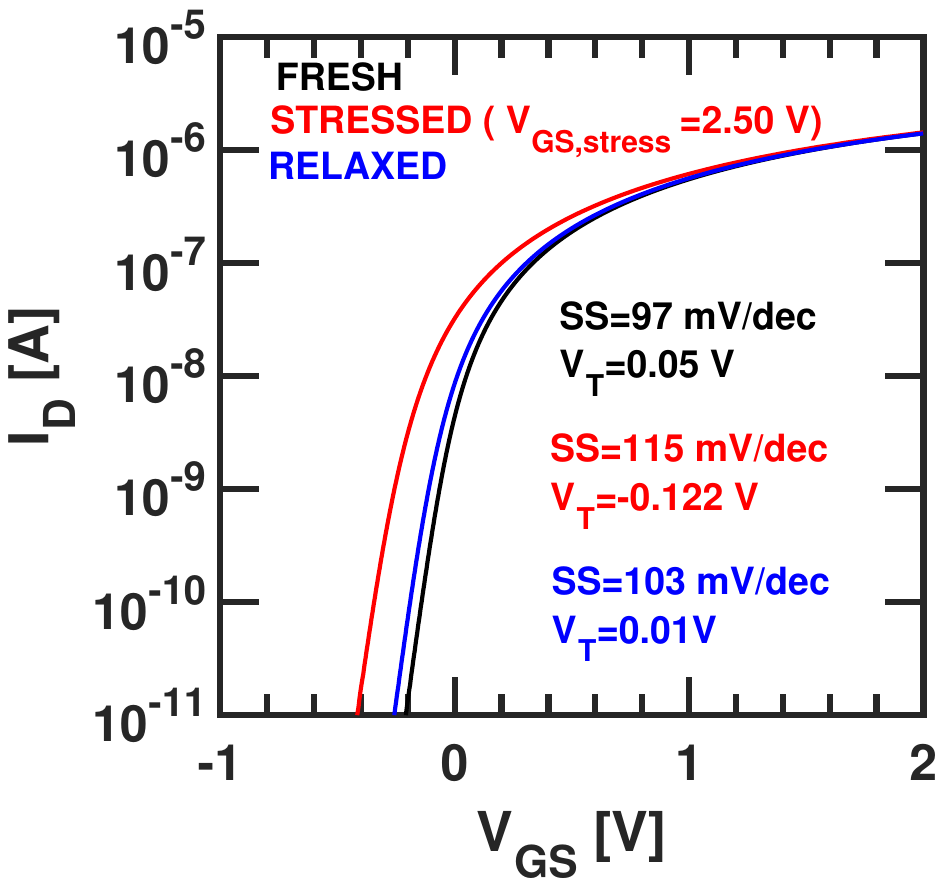}
    \centering
    \includegraphics[width=0.24\textwidth]{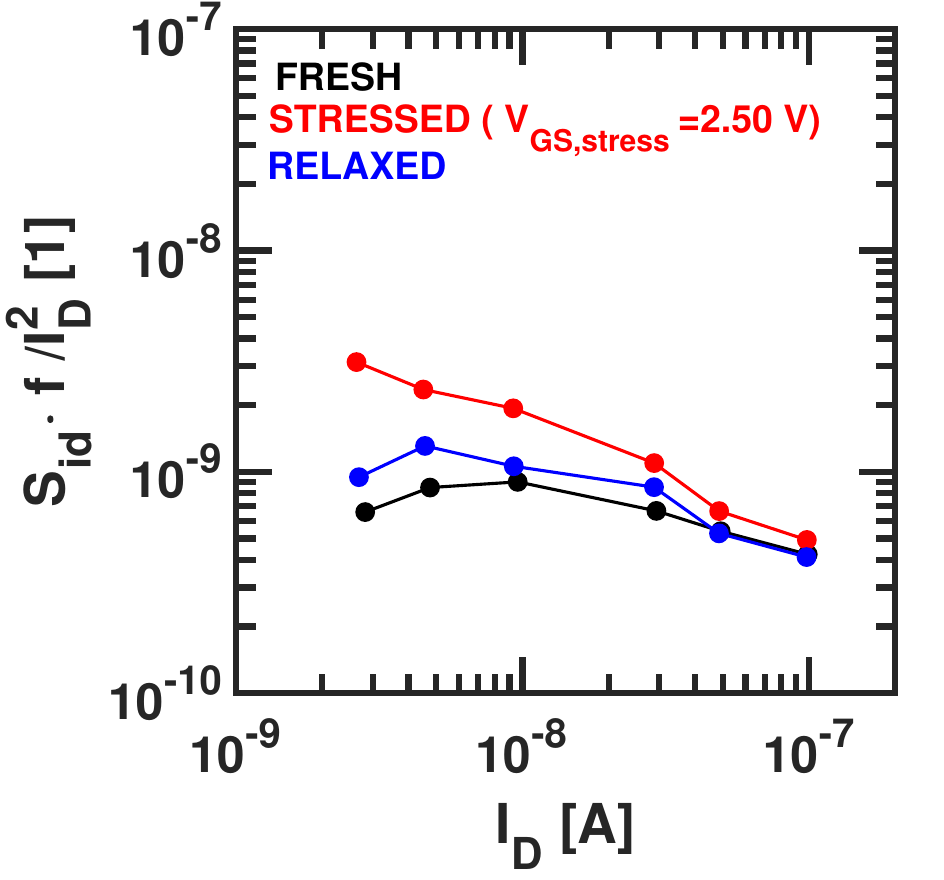} \\ 
    \caption{{Comparison of $I_{\textrm{D}}{-}V_{\textrm{GS}}$ and 1/f noise between fresh, stressed ($V_{\mathrm{GS,stress}}$=2.50 V), and recovered TFT. Stress and recovery were done at $T$=125~°C, while the measurements are done at $T$=25~°C.}}
    \label{fig:Recovery}
\end{figure}

\begin{figure}[]
    \captionsetup[subfigure]{width=0.95\textwidth}
    \centering
    \begin{subfigure}[t]{0.5\columnwidth}
        \centering
        \begin{tikzpicture}
            \node[anchor=south west,inner sep=0] (image) at (0,0) {\includegraphics[width=1\columnwidth]{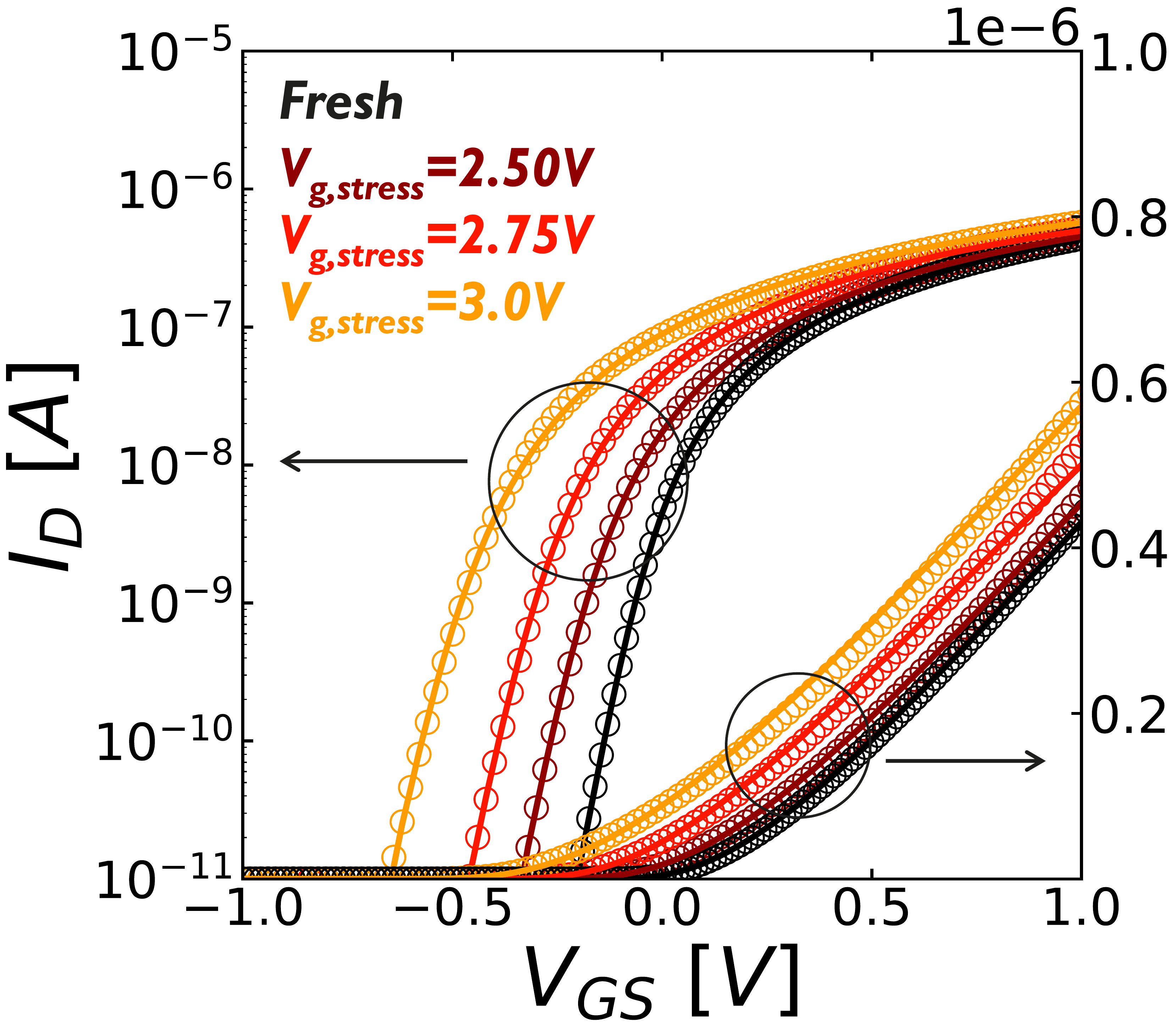}};         
            \node[align=left, anchor=north west] at (0.9,1.2) {(a)};
        \end{tikzpicture}
        \phantomsubcaption
        \label{fig:IV_fitted}
    \end{subfigure}%
    \begin{subfigure}[t]{0.5\columnwidth}
        \centering
        \begin{tikzpicture}
            \node[anchor=south west,inner sep=0] (image) at (0,0) {\includegraphics[width=1\columnwidth]{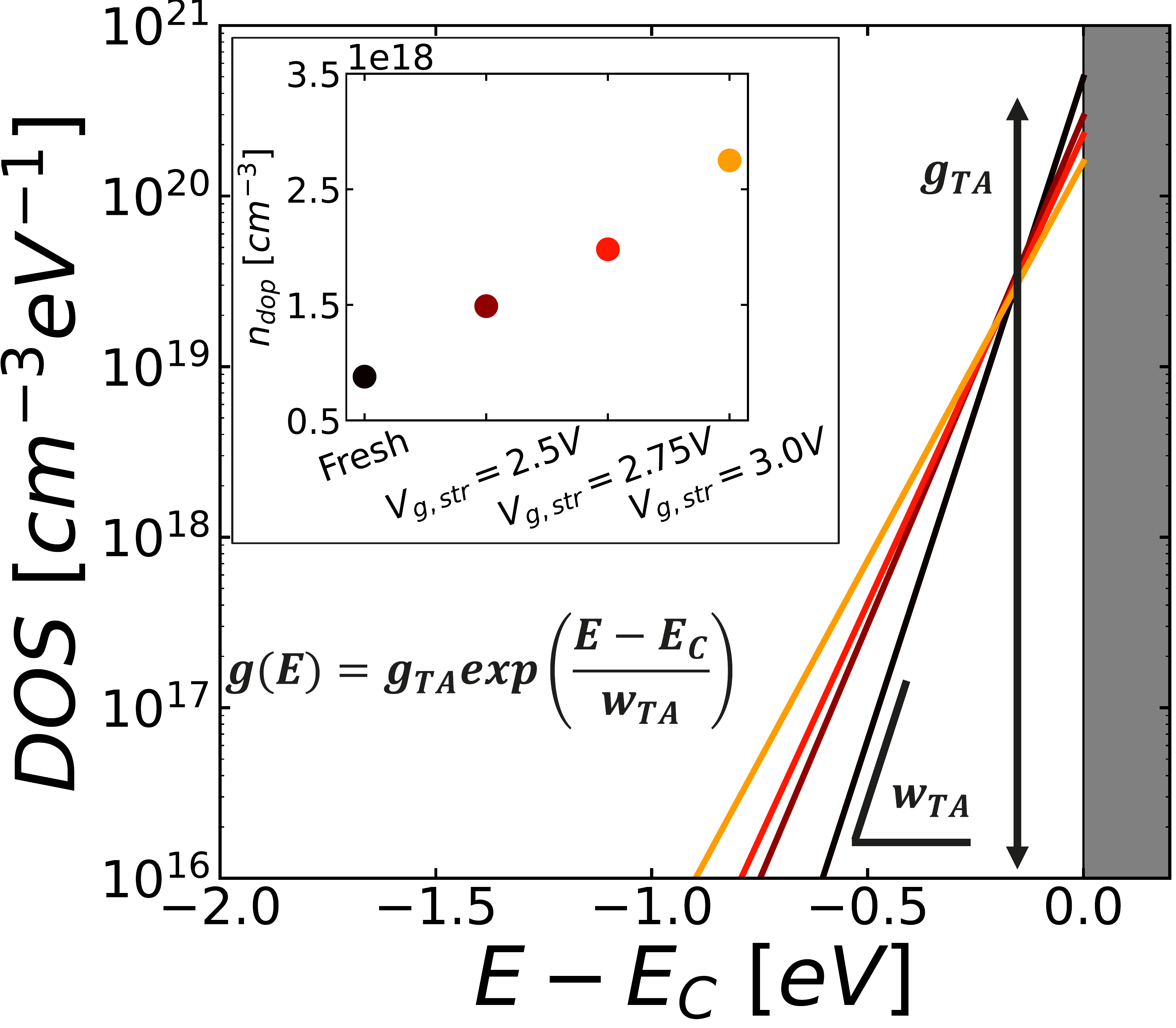}};        
            \node[align=left, anchor=north west] at (0.85,1.15) {(b)};  
        \end{tikzpicture}
        \phantomsubcaption        
        \label{fig:tailEvol}
    \end{subfigure}   
    \caption{(a) Comparison between experimental (circles) and simulated (lines) $I_{\mathrm{D}}-V_{\mathrm{GS}}$. (b) The simulations show a progressive tail broadening with increasing negative shift due to enhanced channel doping (as shown in the inset). This observation aligns with the theoretical Coulomb-induced disorder model of \cite{PhysRevB.63.081202}.}
    \label{fig:TCAD}
\end{figure}
\begin{figure}[]
    \captionsetup[subfigure]{width=0.95\textwidth}
    \centering
    \begin{subfigure}[t]{0.5\columnwidth}
        \centering
        \begin{tikzpicture}
            \node[anchor=south west,inner sep=0] (image) at (0,0) {\includegraphics[width=1\columnwidth]{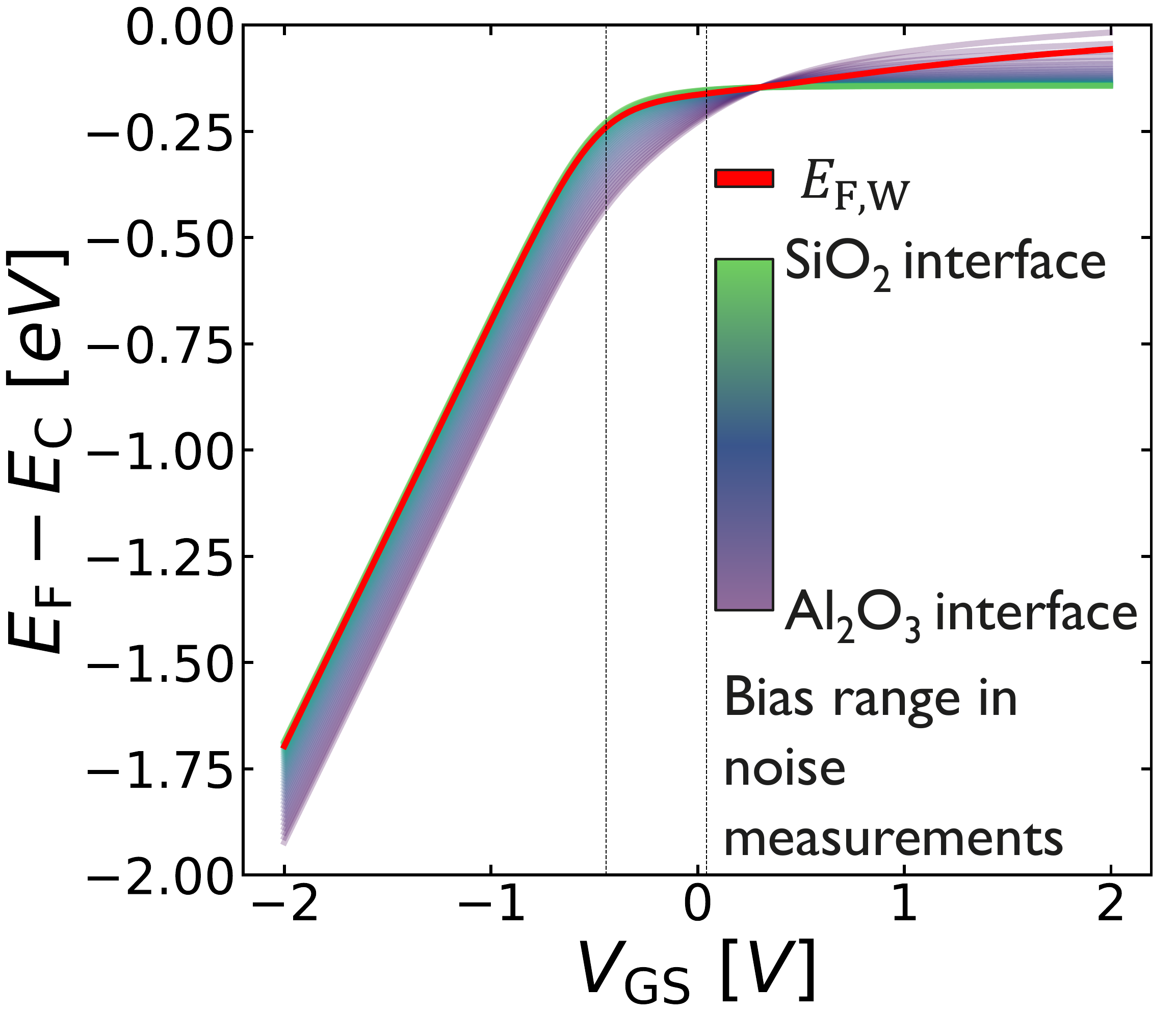}}; 
            \node[align=left, anchor=north west] at (1,3.7) {(a)};
        \end{tikzpicture}
        \phantomsubcaption
        \label{fig:Weighted_Ef}
    \end{subfigure}%
    \begin{subfigure}[t]{0.5\columnwidth}
        \centering
        \begin{tikzpicture}
            \node[anchor=south west,inner sep=0] (image) at (0,0) {\includegraphics[width=1\columnwidth]{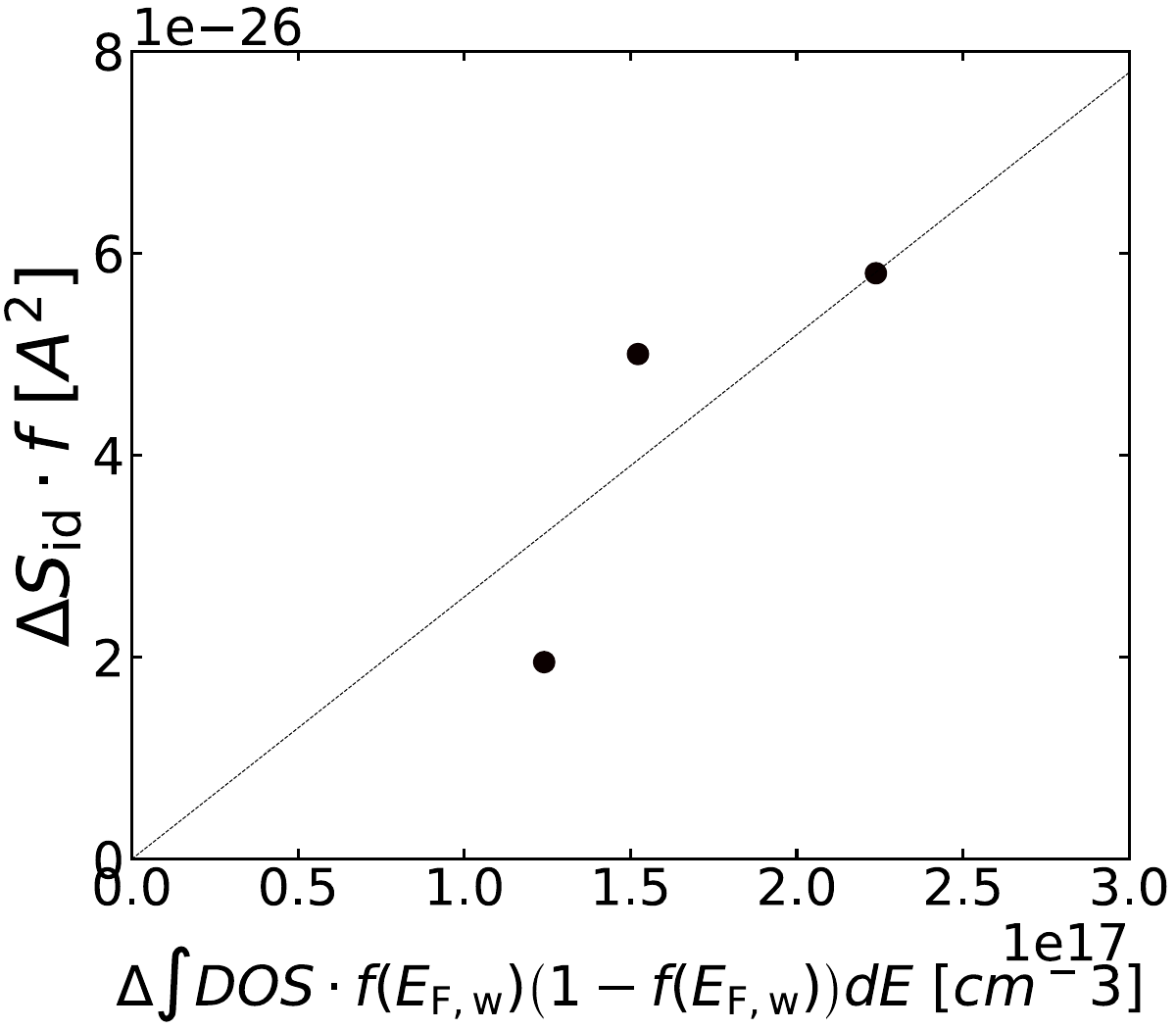}};        
            \node[align=left, anchor=north west] at (0.47,3.7) {(b)};  
        \end{tikzpicture}
        \phantomsubcaption        
        \label{fig:DeltaDos_plot}
    \end{subfigure}   

    \caption{{(a) Simulated Fermi level $E_\mathrm{F}$ (referenced to the conduction band edge $E_\mathrm{C}$) as a function of $V_\mathrm{GS}$ at various distances from the $\mathrm{Al_2O_3}$/channel interface to the channel/$\mathrm{SiO_2}$ interface (see \cref{fig:Schematic}). The red curve represents the weighted average Fermi level ($E_\mathrm{F,W}$), calculated using the local current density as a weighting factor. (b) Numerical correlation between tail charge fluctuations at $I_\mathrm{D} = 2$ nA (calculated as in \cite{10.1063/1.4994152}) and the corresponding change in the measured noise power spectral density ($\Delta S_\mathrm{id}\cdot f$).}}
    \label{fig:DeltaDos_calculations}
\end{figure}

\begin{figure}[!b]
\centering
    \includegraphics[width=0.44\textwidth]
    {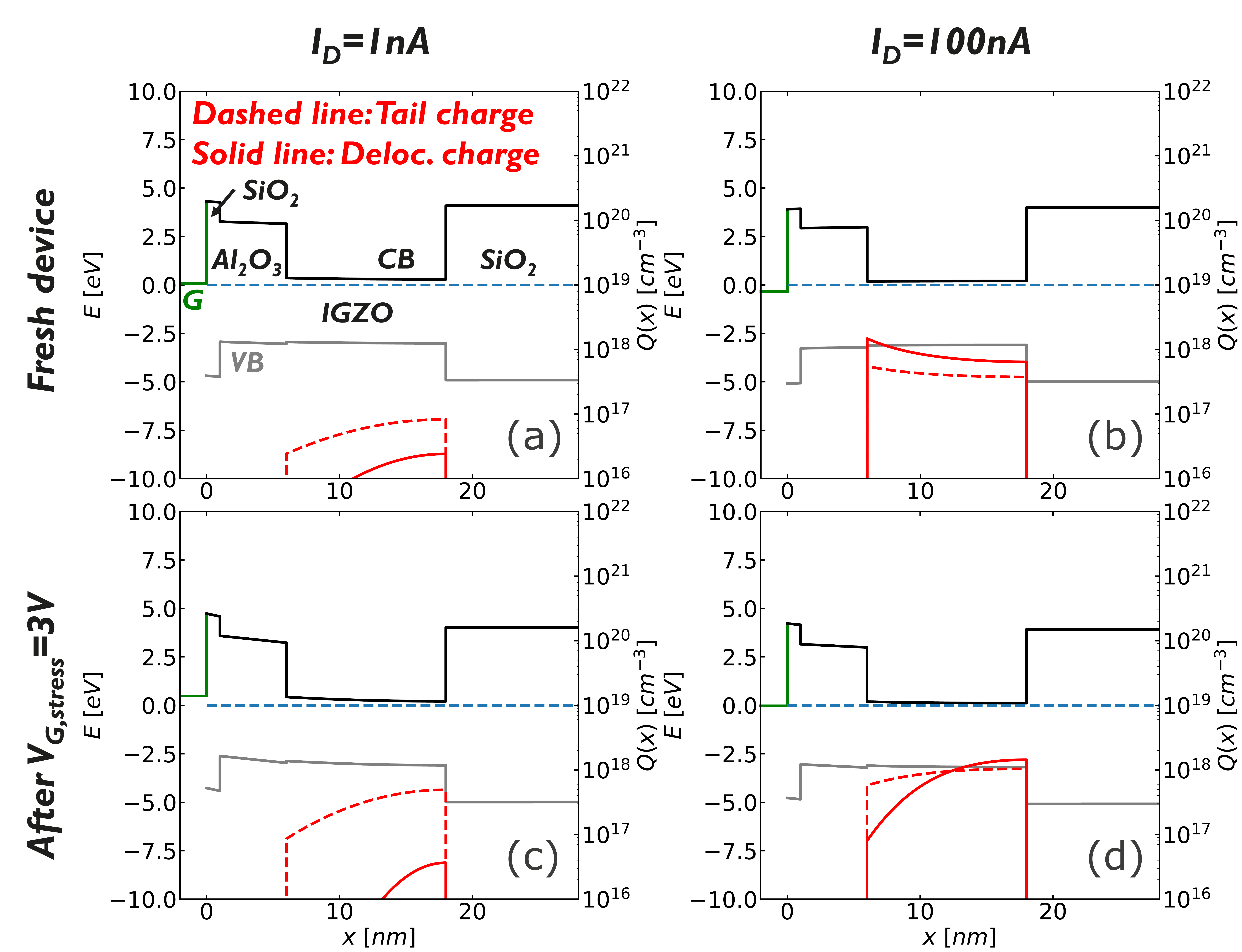} 
    \caption{Band diagram (left y-axis) and charge densities (right y-axis) along the direction perpendicular to the IGZO/dielectric interfaces ($x$) for a fresh TFT and one stressed at $V_{\mathrm{GS,stress}}$=3 V.}
    \label{fig:100nm_tch_electrostatics}
\label{fig:ChargeProfiles}
\end{figure}
To investigate the common cause of SS and noise degradation with BTI stress, we simulate the pre- and post-stress $I_{\mathrm{D}}-V_{\mathrm{GS}}$ using {an in-house Poisson solver \cite{ESolver} that imposes continuity at the transition between conduction band and exponential band tail states \cite{10.1063/1.5008521}, a specific feature of amorphous semiconductors \cite{Davis1970}. The $I_{\mathrm{D}}-V_{\mathrm{GS}}$ fit involves only four IGZO parameters: doping level $n_\mathrm{{dop}}$, peak tail density $g_\mathrm{{TA}}$, characteristic energy $w_\mathrm{{TA}}$
and (constant) mobility $\mathrm{\mu}$}. 
\cref{fig:IV_fitted} shows excellent agreement between experimental and calibrated $I_{\mathrm{D}}-V_{\mathrm{GS}}$ (parameters listed in \cref{tab:params_overview}). Doping increases with stress due to BTI-induced hydrogen diffusion into the channel. Remarkably, the fits confirm a \textit{broadening} of tail states after stress (\cref{fig:tailEvol}), in line with the noise increase seen in \cref{fig:Noise_Stressed}. This increase in disorder induced by doping has been theoretically investigated in \cite{PhysRevB.63.081202}, and our results are in line with their study. {We assume a constant mobility because the solver does not model the complex mobility mechanisms that arise in IGZO after stress and hydrogen incorporation. This simplification is justified by the compensation between the mobility loss due to additional scattering and the mobility gain due to higher carrier concentration from doping \cite{Takagi2005,Kamiya2010}.}
\par {We quantify the correlation between band tail broadening and increase in 1/f noise by evaluating the fluctuation of band tail charge around the Fermi level. Using our in-house simulator, we extract the average Fermi level ($E_\mathrm{F,w}$) across the channel, weighted by the local current density. The band tail charge fluctuation is then computed following the method in \cite{10.1063/1.4994152}. As shown in \cref{fig:DeltaDos_plot}, this fluctuation correlates well with the variation in the measured noise power spectral density ($\Delta S_\mathrm{id} \cdot f$), further supporting that the noise increase is driven by band tail broadening.} 
\par The specific microscopic mechanism causing an increase in noise due to the band tail states broadening upon PBTI remains under debate. The increase in doping due to hydrogen incorporation in the IGZO film leads to additional Coulomb 
interaction in the IGZO  channel, which affects transport (and noise) until it is screened at higher surface potentials. This phenomenon could explain the observed high noise degradation at low $I_{\mathrm{D}}$, while no degradation occurs at higher $I_{\mathrm{D}}$ when this charge is effectively screened by the channel electrons. 
\begin{table}[]
\centering
\resizebox{1\columnwidth}{!}{%
\begin{tabular}{|l|l|l|l|l|l|l|}
\hline
\textbf{Parameter} & \textbf{Unit}                        & \fbox{\textbf{Fresh}}   & \textbf{$\mathbf{V_{G,str}=2.5V}$} & \textbf{$\mathbf{V_{G,str}=2.75V}$} & \textbf{$\mathbf{V_{G,str}=3.0V}$}\\ \hline
\fbox{$n_\mathrm{dop}$}         & \fbox{$\mathrm{[cm^{-3}]}$}          & 8.79$\cdot$10\textsuperscript{17} & 1.49$\cdot$10\textsuperscript{18} & 1.98$\cdot$10\textsuperscript{18} & 2.75$\cdot$10\textsuperscript{18} \\ \hline
\fbox{$g_\mathrm{TA}$}          & \fbox{$\mathrm{[cm^{-3}eV^{-1}]}$}   & 3.00$\cdot$10\textsuperscript{20} & 1.78$\cdot$10\textsuperscript{20} & 1.38$\cdot$10\textsuperscript{20} & 9.60$\cdot$10\textsuperscript{19}  \\ \hline
\fbox{$w_\mathrm{TA}$}          & \fbox{$\mathrm{[eV]}$}               & 0.056   & 0.073 & 0.079 & 0.092  \\ \hline
\fbox{$\mu$}                    & \fbox{$\mathrm{[cm^2V^{-1}s^{-1}]} $}& 13.3      & 13.3   & 13.3 & 13.3   \\ \hline
\end{tabular}%
}
\vspace{1.5pt}
\caption{Simulation parameters used to reproduce $I_{\mathrm{D}}-V_{\mathrm{GS}}$ in \cref{fig:IdVg_Stressed}.}
\label{tab:params_overview}
\end{table}
\par Comparing charge distributions before and after stress helps to rule out the impact of dielectric traps on noise. In the subthreshold region (\cref{fig:ChargeProfiles}a,c), the current flows mainly along the backside of the channel, making noise measurements insensitive to gate dielectric defects. In contrast, in the accumulation region ($I_{\mathrm{D}} = 100$~nA), the current flows near the gate dielectric interface in the fresh device (\cref{fig:ChargeProfiles}b), but shifts toward the backside interface in the stressed device due to increased doping (\cref{fig:ChargeProfiles}d). Notably, the noise at $I_{\mathrm{D}} = 100$~nA is nearly identical for both devices (\cref{fig:Noise_Stressed}), despite current flowing through different IGZO/dielectric interfaces, further indicating that the noise originates from the channel.
\section{Conclusions}
\label{sec:conclusions}
We studied high-$T$ PBTI stress in back-gated IGZO devices using DC measurements, 1/f noise, and physics-based simulations to understand the origin of negative $\Delta V_{\textrm{T}}$. The 1/f noise of fresh devices in subthreshold follows the mobility-fluctuation model, suggesting that the noise originates from the channel rather than the dielectric. Noise and SS both increase with PBTI stresses at higher $V_\mathrm{GS}$ and temperature, indicating an increase of conduction band tail states. {A recovery experiment shows that the $V_\mathrm{T}$, SS, and noise values return to the fresh ones after relaxation, confirming that the process is reversible and further supporting the hypothesis of hydrogen exchanges between the dielectric and the channel.} Physics‑based simulations calibrated to the experimental data confirm a correlation between hydrogen‑induced doping and tail‑state broadening. These findings link PBTI at high temperatures to DOS variations, \textit{in addition to} trapping in the gate dielectric.

\clearpage
\bibliographystyle{IEEEtran}
\bibliography{EDL}
\end{document}